\documentclass[aps,twocolumn,floatfix]{revtex4-1}

\usepackage{graphicx}
\usepackage{dcolumn}
\usepackage{bm}
\usepackage{epsf}
\usepackage{subfigure}
\usepackage{epstopdf}
\usepackage{amsmath}
\usepackage{amssymb}

\begin{document}

\title{Nonequilibrium thermodynamics and Nose-Hoover dynamics}
\author{Massimiliano Esposito}%
\email{mesposit@ulb.ac.be}
\affiliation{Center for Nonlinear Phenomena and Complex Systems, Universit\'e Libre de Bruxelles Campus Plaine CP231 B-1050 Brussels, Belgium}
\author{Takaaki Monnai}%
\email{monnai@suou.waseda.jp}%
\affiliation{Department of Applied Physics, Osaka City University, 3-3-138 Sugimoto, Sumiyoshi-ku, Osaka 558-8585, Japan}

\date{\today}

\begin{abstract}
We show that systems driven by an external force and described by Nose-Hoover dynamics allow for a consistent nonequilibrium thermodynamics description when the thermostatted variable is initially assumed in a state of canonical equilibrium. By treating the ``real" variables as the system and the thermostatted variable as the reservoir, we establish the first and second law of thermodynamics. As for Hamiltonian systems, the entropy production can be expressed as a relative entropy measuring the system-reservoir correlations established during the dynamics. 
\end{abstract}

\maketitle

\section{Introduction}

The thermodynamic description of a system out of equilibrium is based on the first and second law. The first law states that since energy is conserved, the change in the system energy is the sum of the energy added doing work on the system and the energy flowing into the system from the environment under the form of heat. The second law states that the change in system entropy is the sum of the entropy flow, a reversible term given by heat divided by temperature, and the entropy production, an always positive or zero irreversible term \cite{PrigoThermo,Prigogine,GrootMazur}. During the last years, significant progress has been achieved in establishing the relation between the nonequilibrium thermodynamics description of a system and its underlying dynamics. The discovery of fluctuation theorems, first for thermostatted deterministic dynamics \cite{Evans93, Evans94, Gallavotti95, Gallavotti95b, SearlesEvans00, EvansMorrisB} and then for stochastic \cite{Kurchan98, Lebowitz99, Crooks99, Crooks00, Seifert05, EspositoVdBPRL10} and Hamiltonian dynamics \cite{Jarzynski97, Jarzynski97b, HTasaki00, Maes03, Mukamel03, Monnai05, EspositoReview, AndrieuxGaspardTasaki}, played an important role in this regard. For stochastic dynamics, this connection is nowadays well established and has given rise to the field of stochastic thermodynamics \cite{Crooks98, SeifertST08, Harris07, EspoVdB10_Da, EspoVdB10_Db}. More recently, exact relations for the entropy production have also been obtained for Hamiltonian dynamics when considering specific class of initial conditions \cite{VandenBroeck07, Parrondo2009, JarzynskiEPL09, EspoLindVdBNJP10}. Some work in this direction has been done for thermostatted deterministic dynamics \cite{Gallavotti08,Ruelle03,Holian}. 

In this paper, we explicitly construct the thermodynamics description of a driven system with an underlying Nose-Hoover thermostatted deterministic dynamics \cite{Nose84,Hoover85,EvansMorrisB}. Our result can be viewed as the analogue for Nose-Hoover dynamics of the recent result obtained in \cite{EspoLindVdBNJP10} for open systems described by Hamiltonian dynamics. A key point in establish this connection is to treat the thermostatted variable as the reservoir. Our sole assumption is that the initial probability distribution of the thermostatted variable be a canonical equilibrium one. 

In section \ref{NHDyn} we briefly remind the key properties of Nose-Hoover dynamics. In section \ref{SysBathSep} we operate the system-reservoir identification and in section \ref{FirstlawSec} we identify heat and work to establish the first law of thermodynamics. In section \ref{SeclawSec} we identify entropy and entropy production to establish the second law of thermodynamics. In section \ref{Irrwork} we discuss the connection between irreversible work and entropy production and in section \ref{CorrEnt} we discuss the interpretation of entropy production in term of system-reservoir correlations. Conclusions are drawn in section \ref{Conc}. 

\section{Nose-Hoover dynamics}\label{NHDyn}

We consider a $N$-particle system confined in a $D$ dimensional time dependent potential $V(\lambda(t),\{q_i(t)\})$ in contact with a Nose-Hoover thermostat at temperature $T$ \cite{Nose84,Hoover85,EvansMorrisB}. The coordinates and conjugate momenta of the particles are denoted by $q_i$ and $p_i (1\leq i\leq DN)$ and the time dependence of the potential occurs through the driving parameter $\lambda$. The equations of motion read:
\begin{eqnarray}
&&\dot{q}_i(t) = \frac{p_i(t)}{m_i} \nonumber \\
&&\dot{p}_i(t) = -\frac{\partial V(\lambda(t),\{q_i(t)\})}{\partial q_i(t)}-\zeta(t) p_i(t) \nonumber \\
&&\dot{\zeta}(t) = \frac{1}{\alpha} \big( \sum_{i=1}^{DN} \frac{p_i^2}{m_i} - D N k_b T \big). \label{EQM}
\end{eqnarray}
The dynamical variable $\zeta$ mimics a friction coefficient and $\alpha$ is a measure of the relaxation rate. Hereafter, we abbreviate the set of variables $\{q_i,p_i,\zeta\}$ and $\{q_i,p_i\}$ by $\Gamma$ and $\Gamma_s$ respectively.  

Any probability distribution $f(t,\Gamma(t))$ on the phase space $\Gamma$, due to conservation, satisfies
\begin{eqnarray}
\frac{\partial f}{\partial t} = - \frac{\partial}{\partial \Gamma} (\dot{\Gamma} f) 
= - (\frac{\partial \dot{\Gamma}}{\partial \Gamma}) f - \frac{\partial f}{\partial \Gamma} \dot{\Gamma} \label{ProbEvol1}
\end{eqnarray}
As a result, from 
\begin{eqnarray}
\frac{d f}{dt} = \frac{\partial f}{\partial t} + \frac{\partial f}{\partial \Gamma} \dot{\Gamma}, \label{ProbEvol2}
\end{eqnarray}
we get that
\begin{eqnarray}
\frac{d f}{d t} = - (\frac{\partial \dot{\Gamma}}{\partial \Gamma}) f = \frac{d \ln f}{dt} f = - \Lambda f \label{ProbEvol3}
\end{eqnarray}
where $\Lambda$ is the phase space compression factor. For Hamiltonian dynamics, Liouville theorem implies $\Lambda=0$, but for the Nose-Hoover dynamics (\ref{EQM}), the phase space compression factor is given by $\Lambda = - DN \zeta(t)$. This means that any probability distribution evolves in time according to  
\begin{equation}
f(t,\Gamma(t)) = e^{DN \int_0^t ds \zeta(s)} f(0,\Gamma(0)) \label{ProbEvol4}
\end{equation}
which also implies 
\begin{equation}
d\Gamma(t)=d\Gamma(0) e^{- DN \int_0^t ds \zeta(s)}. \label{DeltaEvol}
\end{equation}
These two relations will often be used in what follows.

\section{System bath separation} \label{SysBathSep}

We are now going to consider that the $\Gamma_s$ variables constitute the system, while $\zeta$ constitutes the reservoir. This is reasonable since $\Gamma_s$ are the true physical variables while $\zeta$ is only an artificial variable introduced to mimic the effect of a reservoir on the system dynamics. As a result, we define the (reduced) system and reservoir probability distributions as
\begin{eqnarray}
&& f_s(t,\Gamma_s(t)) = \int d\zeta(t) f(t,\Gamma(t)) \\
&& f_r(t,\zeta(t)) = \int d\Gamma_s(t) f(t,\Gamma(t)) \label{RedDensity} 
\end{eqnarray}
and the system and reservoir Hamiltonian as
\begin{eqnarray}
&& H_s(\lambda,\Gamma_s) = \sum_{i=1}^{DN} \frac{p_i^2}{2m_i} + V(\lambda,\{q_i\}) \nonumber \\
&& H_r(\zeta) = \frac{\alpha \zeta^2}{2} . \label{HsysRes}
\end{eqnarray}
The system and reservoir canonical distributions are therefore defined as 
\begin{eqnarray}
f^{eq}_s(\lambda,\Gamma_s) = \frac{e^{-\beta H_s(\lambda,\Gamma_s)}}{Z_s(\lambda)} \ \ , \ \ f^{eq}_r(\zeta) = \frac{e^{-\beta H_r(\zeta)}}{Z_r}.  \label{CanDistribSys}
\end{eqnarray}
Note that $\beta=(k_b T)^{-1}$ and that $Z_r$ and $Z_s(\lambda)$ are the system and reservoir partition functions
\begin{eqnarray}
Z_s(\lambda) = \int d\Gamma_s e^{-\beta H_s(\lambda,\Gamma_s)} \ \ , \ \ Z_r = \int d\zeta e^{-\beta H_r} .\label{PartFct}
\end{eqnarray}  
We finally introduce the canonical distribution of the total system 
\begin{equation}
f^{eq}(\lambda,\Gamma) = \frac{e^{-\beta H_0(\lambda,\Gamma)}}{Z_r Z_s(\lambda)}=f^{eq}_s(\lambda,\Gamma_s) f^{eq}_r(\zeta) , \label{CanDistrib}
\end{equation}
associated to the system-reservoir Hamiltonian
\begin{eqnarray}
H_0(\lambda,\Gamma) = H_s(\lambda,\Gamma_s) + H_r(\zeta). \label{Htot}
\end{eqnarray}
An important property of the canonical distribution (\ref{CanDistrib}) is that it is invariant under the Nose-Hoover dynamics for constant value of the driving parameter $\lambda$. Indeed, using (\ref{EQM}), we verify that $\dot{\Gamma} \frac{\partial f^{eq}}{\partial \Gamma}=-\Lambda f^{eq}$ and that, using (\ref{ProbEvol2}) and (\ref{ProbEvol3}), $\frac{\partial f^{eq}}{\partial t}=0$.

\section{Work and heat} \label{FirstlawSec}

We start by noting that under the Nose-Hoover dynamics (\ref{EQM}), the Hamiltonian (\ref{Htot}) evolves according to 
\begin{eqnarray}
\frac{d H_{0}(\lambda(t),\Gamma(t))}{dt} = \frac{\partial H_s(\lambda(t),\Gamma_s(t))}{\partial \lambda(t)} \dot{\lambda}(t) - DN k_b T \zeta(t) . \nonumber\\ \label{HextEv}
\end{eqnarray}
This shows that even in absence of external driving, i.e. for a fixed value of $\lambda$, the Hamiltonian (\ref{EQM}) is not conserved. Since the work performed on the system by the external driving from $0$ to $t$ is naturally defined as 
\begin{eqnarray}
W[t,\Gamma_s] = \int_{0}^{t} d\tau \frac{\partial H_s(\lambda(\tau),\Gamma_s(\tau))}{\partial \lambda(\tau)} \dot{\lambda}(\tau) \label{Work},
\end{eqnarray}
using (\ref{HextEv}), we can express this work as
\begin{eqnarray}
W[t,\Gamma_s] &=& \Delta H_{s}[t,\Gamma_s] + \Delta H_r[t,\zeta] \nonumber\\
&& + DN k_b T \int_{0}^{t} ds \zeta(s) , \label{DeltaDef}
\end{eqnarray}
where the energy change in the system and reservoir along the trajectory $\Gamma$ reads
\begin{eqnarray}
&& \Delta H_s[t,\Gamma_s] = H_{s}(\lambda(t),\Gamma_s(t)) - H_{s}(\lambda(0),\Gamma_s(0)) \nonumber\\
&& \Delta H_r[t,\zeta] = H_r(\zeta_t)-H_r(\zeta_0) \label{DeltaH} .
\end{eqnarray}
Since the first law of thermodynamics should apply for the system energy
\begin{eqnarray}
\Delta H_s[t,\Gamma_s] = W[t,\Gamma_s] + Q[t,\zeta] , \label{FirstP}
\end{eqnarray}
it becomes natural to define heat as
\begin{eqnarray}
Q[t,\zeta] = -\Delta H_r[t,\zeta] - DN k_b T \int_{0}^{t}d\tau \zeta(\tau) .\label{DefHeat}
\end{eqnarray}

In the following, the statistical average of a trajectory dependent quantity $X[t,\Gamma(t)]$ will be denoted by
\begin{eqnarray}
X(t) &=& \langle X[t,\Gamma(t)] \rangle_t \nonumber \\
&=& \int d\Gamma(t) f(t,\Gamma(t)) X[t,\Gamma(t)]. \label{aver}
\end{eqnarray}

\section{Entropy and entropy production} \label{SeclawSec}

Our key assumption is that we are going to consider initial conditions of the form
\begin{eqnarray}
f(0,\Gamma(0)) = f_s(0,\Gamma_s(0)) f^{eq}_r(\zeta(0)) \label{Inicond},
\end{eqnarray}
where $f_s(0,\Gamma_s(0))$ is an arbitrary system distribution and $f^{eq}_r(\zeta(0))$ is the the canonical equilibrium distribution of the reservoir. 

The central result of this paper is that the entropy production is given by
\begin{eqnarray}
\Delta_i S(t) = k_b \bigg\langle \ln \frac{f(t,\Gamma(t))}{f_s(t,\Gamma_s(t)) f_r^{eq}(\zeta(t))} \bigg\rangle_t. \label{EPexpl}
\end{eqnarray}
Indeed, using (\ref{ProbEvol4}) and (\ref{DefHeat}), we can express (\ref{EPexpl}) as
\begin{eqnarray}
\Delta_i S(t) = \Delta S(t) - \Delta_e S(t), \label{2ndLaw}
\end{eqnarray}
where the first term $\Delta S(t) = S(t) - S(0)$ is the change in the system entropy expressed as the change in Shannon entropy associated to the reduced system probability distribution
\begin{equation} 
S(t) = - k_b \int d\Gamma_s(t) f_s(t,\Gamma_s(t)) \ln f_s(t,\Gamma_s(t)) , \label{Shannon} 
\end{equation}
and the second term is the entropy flow expressed as the heat (\ref{DefHeat}) divided by temperature 
\begin{eqnarray}
\Delta_e S(t) = \frac{Q(t)}{T}. \label{EntropyFlow}
\end{eqnarray}
The entropy production (\ref{2ndLaw}) is always positive or zero since it can be rewritten as a relative entropy 
\begin{eqnarray}
\Delta_i S(t) = k_b {\rm D} \big[ f(t,\Gamma(t)) \big| \big| f_s(t,\Gamma_s(t)) f_r^{eq}(\zeta(t)) \big] \geq 0 . \nonumber \\ \label{EP}
\end{eqnarray}
Properties of relative entropies are discussed in detail in \cite{CoverThomas}. The entropy production will be zero only when $f(t,\Gamma(t)) = f_s(t,\Gamma_s(t)) f_r^{eq}(\zeta(t))$. It thus becomes clear that (\ref{2ndLaw}) constitutes the second law of thermodynamics stating that the change in the system entropy $\Delta S(t)$ is the sum of two contributions, the irreversible entropy production $\Delta_i S(t)\geq 0$ and the reversible entropy flow due to heat exchanges $\Delta_e S(t)$ \cite{PrigoThermo,Prigogine,GrootMazur}. Because a relative entropy is a measure of how different two distributions are from each other, we get a appealing physical interpretation of entropy production as a measure of how different the actual distribution $f(t,\Gamma(t))$ is from the product of the actual reduced probability distribution of the system and the equilibrium distribution of the reservoir. The same result has been found for Hamiltonian dynamics in \cite{EspoLindVdBNJP10}. Introducing the nonequilibrium free energy 
\begin{eqnarray}
F(t) = H_s(t) - T S(t) \label{FreeE}
\end{eqnarray}
and using (\ref{2ndLaw}), (\ref{EntropyFlow}) and the average of (\ref{FirstP}), we can also express entropy production as 
\begin{eqnarray}
\Delta_i S(t) = \frac{W(t) - \Delta F(t)}{k_b T} \geq 0 \label{EP2}.
\end{eqnarray}

\section{Irreversible work} \label{Irrwork} 

We now consider the special case where the initial probability distribution of the system (\ref{Inicond}) is the canonical equilibrium distribution (\ref{CanDistribSys}). In this case it is interesting to compare the entropy production (\ref{EP2}) with the irreversible work defined as
\begin{eqnarray}
W_{diss}(t) = \frac{W(t) - \Delta F^{eq}(t)}{k_b T} , \label{EPSimple} 
\end{eqnarray}
where $\Delta F^{eq}(t)=F^{eq}(t)-F^{eq}(0)$ is the change in the equilibrium free energy $F^{eq}(t)=-\beta^{-1} \ln Z_s(\lambda(t))$. Indeed, using (\ref{ProbEvol4}) and (\ref{DeltaDef}), we find that the irreversible work can be expressed as 
\begin{eqnarray}
W_{diss}(t) = k_b {\rm D} \big[ f(t,\Gamma(t)) \big| \big| f^{eq}(\lambda(t),\Gamma(t)) \big] \geq 0 , \label{EPSimple2}
\end{eqnarray}
and is thus a direct measure of the lag between the actual and the canonical probability distribution in the join system-reservoir space. The same result was obtained in \cite{JarzynskiEPL09} for Hamiltonian and stochastic dynamics. Closely related results have also been obtained for Hamiltonian dynamics in \cite{VandenBroeck07,Parrondo2009}. We can now express the difference between irreversible work and entropy production as a relative entropy between the actual and the equilibrium system distribution 
\begin{eqnarray}
W_{diss}(t) - \Delta_i S(t) = k_b {\rm D} \big[ f_s(t,\Gamma_s(t)) \big| \big| f^{eq}_s(\lambda(t),\Gamma_s(t)) \big] . \nonumber \\ \label{Diff}
\end{eqnarray}
Due to the positivity of relative entropies we get the inequality
\begin{eqnarray}
W_{diss}(t) \geq \Delta_i S(t) \geq 0, \label{Inequal1}
\end{eqnarray}
which only becomes an equality when the final state of the system (at time $t$) is the canonical equilibrium distribution. Using (\ref{EP2}) with (\ref{EPSimple2}), we also find that the equilibrium free energy is the minimum of the nonequilibrium free energy
\begin{eqnarray}
F(t) \geq F^{eq}(t) . \label{Inequal2}
\end{eqnarray}

\section{Correlation entropy}\label{CorrEnt}

Using (\ref{ProbEvol4}), we start by noting that the change in the total Shannon entropy 
\begin{equation} 
S_{tot}(t) = - k_b \int d\Gamma(t) f(t,\Gamma(t)) \ln f(t,\Gamma(t)) \label{ShannonTot} 
\end{equation}
is given by \cite{Holian,Evans85}
\begin{eqnarray}
\Delta S_{tot}(t) = - k_b DN \int_0^t ds \langle \zeta(s) \rangle_s . \label{ChShannonTot}
\end{eqnarray}
We define the correlation entropy as minus the mutual system-reservoir information, i.e. the difference between the Shannon entropy of the total system $S_{tot}(t)$ and the sum of the Shannon entropy of the system $S(t)$ and of the reservoir $S_r(t)$
\begin{eqnarray}
S_{c}(t) = S_{tot}(t)-S(t)-S_r(t) \label{CorrEnt1}.
\end{eqnarray}
For the initial conditions that we consider (\ref{Inicond}), $S_c(0)=0$ and the correlation entropy at time $t$ can be expressed as the relative entropy
\begin{eqnarray}
\hspace{-0.4cm} -S_{c}(t) = k_b {\rm D} \big[ f(t,\Gamma(t)) \big| \big| f_s(t,\Gamma_s(t)) f_r(\zeta(t)) \big] \geq 0  \label{CorrEnt2}.
\end{eqnarray}
As suggested by its name, the correlation entropy measures the amount of negative entropy stored in the system-reservoir correlations by the dynamics. The correlation entropy is related to the entropy production by
\begin{eqnarray}
\hspace{-0.4cm} S_{c}(t) + \Delta_i S(t) =  k_b {\rm D} \big[ f_r(t,\zeta(t)) \big| \big| f_r^{eq}(t,\zeta(t)) \big] \geq 0 \label{CorrEnt3},
\end{eqnarray}
which implies the inequality
\begin{eqnarray}
\Delta_i S(t) \geq -S_{c}(t) \geq 0.
\end{eqnarray}
The equality is satisfied when the reservoir can be assumed at equilibrium at time $t$. In such case the entropy production can be interpreted as minus the correlation entropy, i.e. as the mutual system-reservoir information.

\section{Conclusions} \label{Conc}

We have shown in this paper that Nose-Hoover dynamics can be made fully consistent with thermodynamics for class of initial conditions of the kind (\ref{Inicond}). We identified the microscopic expressions for heat, work, system entropy and entropy production and where able to establish the first and second law of thermodynamics starting from the underlying dynamics. This work can be viewed as the analog for thermostatted deterministic dynamics of the recent results obtained for Hamiltonian dynamics in \cite{EspoLindVdBNJP10}. Our microscopic expression for entropy production in term of a relative entropy is reminiscent of results such as \cite{HTasaki00,Maes03,Gaspard04b,VandenBroeck07,EspositoReview}.

\section*{Acknowledgments}

M. E. is supported by the Belgian Federal Government (IAP project ``NOSY"). T. M. owes much to the financial support from JSPS research fellowship for
young scientists.



\begin{thebibliography}{10}%
\makeatletter
\providecommand \@ifxundefined [1]{%
 \ifx #1\undefined \expandafter \@firstoftwo
 \else \expandafter \@secondoftwo
\fi
}%
\providecommand \@ifnum [1]{%
 \ifnum #1\expandafter \@firstoftwo
 \else \expandafter \@secondoftwo
\fi
}%
\providecommand \enquote [1]{``#1''}%
\providecommand \bibnamefont  [1]{#1}%
\providecommand \bibfnamefont [1]{#1}%
\providecommand \citenamefont [1]{#1}%
\providecommand\href[0]{\@sanitize\@href}%
\providecommand\@href[1]{\endgroup\@@startlink{#1}\endgroup\@@href}%
\providecommand\@@href[1]{#1\@@endlink}%
\providecommand \@sanitize [0]{\begingroup\catcode`\&12\catcode`\#12\relax}%
\@ifxundefined \pdfoutput {\@firstoftwo}{%
 \@ifnum{\z@=\pdfoutput}{\@firstoftwo}{\@secondoftwo}%
}{%
 \providecommand\@@startlink[1]{\leavevmode}%
 \providecommand\@@endlink[0]{}%
}{%
 \providecommand\@@startlink[1]{%
  \leavevmode
  \pdfstartlink
   attr{/Border[0 0 1 ]/H/I/C[0 1 1]}%
   user{/Subtype/Link/A<</Type/Action/S/URI/URI(#1)>>}%
  \relax
 }%
 \providecommand\@@endlink[0]{\pdfendlink}%
}%
\providecommand \url  [0]{\begingroup\@sanitize \@url }%
\providecommand \@url [1]{\endgroup\@href {#1}{\urlprefix}}%
\providecommand \urlprefix [0]{URL }%
\providecommand \Eprint[0]{\href }%
\@ifxundefined \urlstyle {%
  \providecommand \doi [1]{doi:\discretionary{}{}{}#1}%
}{%
  \providecommand \doi [0]{doi:\discretionary{}{}{}\begingroup
  \urlstyle{rm}\Url }%
}%
\providecommand \doibase [0]{http://dx.doi.org/}%
\providecommand \Doi[1]{\href{\doibase#1}}%
\providecommand \bibAnnote [3]{%
  \BibitemShut{#1}%
  \begin{quotation}\noindent
    \textsc{Key:}\ #2\\\textsc{Annotation:}\ #3%
  \end{quotation}%
}%
\providecommand \bibAnnoteFile [2]{%
  \IfFileExists{#2}{\bibAnnote {#1} {#2} {\input{#2}}}{}%
}%
\providecommand \typeout [0]{\immediate \write \m@ne }%
\providecommand \selectlanguage [0]{\@gobble}%
\providecommand \bibinfo [0]{\@secondoftwo}%
\providecommand \bibfield [0]{\@secondoftwo}%
\providecommand \translation [1]{[#1]}%
\providecommand \BibitemOpen[0]{}%
\providecommand \bibitemStop [0]{}%
\providecommand \bibitemNoStop [0]{.\EOS\space}%
\providecommand \EOS [0]{\spacefactor3000\relax}%
\providecommand \BibitemShut [1]{\csname bibitem#1\endcsname}%
\bibitem{PrigoThermo}%
  \BibitemOpen
  \bibfield{author}{%
  \bibinfo {author} {\bibfnamefont{I.}~\bibnamefont{Prigogine}},\ }%
  \emph{\bibinfo {title} {Thermodynamics of Irreversible Processes}}\ (\bibinfo
  {publisher} {Wiley-Interscience},\ \bibinfo {year} {1961})%
  \bibAnnoteFile{NoStop}{PrigoThermo}%
\bibitem{Prigogine}%
  \BibitemOpen
  \bibfield{author}{%
  \bibinfo {author} {\bibfnamefont{D.}~\bibnamefont{Kondepudi}}\ and\ \bibinfo
  {author} {\bibfnamefont{I.}~\bibnamefont{Prigogine}},\ }%
  \emph{\bibinfo {title} {Modern thermodynamics}}\ (\bibinfo {publisher}
  {Wiley},\ \bibinfo {year} {1998})%
  \bibAnnoteFile{NoStop}{Prigogine}%
\bibitem{GrootMazur}%
  \BibitemOpen
  \bibfield{author}{%
  \bibinfo {author} {\bibfnamefont{S.~R.}\ \bibnamefont{de~Groot}}\ and\
  \bibinfo {author} {\bibfnamefont{P.}~\bibnamefont{Mazur}},\ }%
  \emph{\bibinfo {title} {Non-equilibrium thermodynamics}}\ (\bibinfo
  {publisher} {Dover},\ \bibinfo {year} {1984})%
  \bibAnnoteFile{NoStop}{GrootMazur}%
\bibitem{Evans93}%
  \BibitemOpen
  \bibfield{author}{%
  \bibinfo {author} {\bibfnamefont{D.~J.}\ \bibnamefont{Evans}}, \bibinfo
  {author} {\bibfnamefont{E.~G.~D.}\ \bibnamefont{Cohen}},\ and\ \bibinfo
  {author} {\bibfnamefont{G.~P.}\ \bibnamefont{Morriss}},\ }%
  \bibfield{journal}{%
  \bibinfo {journal} {Phys. Rev. Lett.}\ }%
  \textbf{\bibinfo {volume} {71}},\ \bibinfo {pages} {2401} (\bibinfo {year}
  {1993})%
  \bibAnnoteFile{NoStop}{Evans93}%
\bibitem{Evans94}%
  \BibitemOpen
  \bibfield{author}{%
  \bibinfo {author} {\bibfnamefont{D.~J.}\ \bibnamefont{Evans}}\ and\ \bibinfo
  {author} {\bibfnamefont{D.~J.}\ \bibnamefont{Searles}},\ }%
  \bibfield{journal}{%
  \bibinfo {journal} {Phys. Rev. E}\ }%
  \textbf{\bibinfo {volume} {50}},\ \bibinfo {pages} {1645} (\bibinfo {year}
  {1994})%
  \bibAnnoteFile{NoStop}{Evans94}%
\bibitem{Gallavotti95}%
  \BibitemOpen
  \bibfield{author}{%
  \bibinfo {author} {\bibfnamefont{G.}~\bibnamefont{Gallavotti}}\ and\ \bibinfo
  {author} {\bibfnamefont{E.~G.~D.}\ \bibnamefont{Cohen}},\ }%
  \bibfield{journal}{%
  \bibinfo {journal} {Phys. Rev. Lett.}\ }%
  \textbf{\bibinfo {volume} {74}},\ \bibinfo {pages} {2694} (\bibinfo {year}
  {1995})%
  \bibAnnoteFile{NoStop}{Gallavotti95}%
\bibitem{Gallavotti95b}%
  \BibitemOpen
  \bibfield{author}{%
  \bibinfo {author} {\bibfnamefont{G.}~\bibnamefont{Gallavotti}}\ and\ \bibinfo
  {author} {\bibfnamefont{E.~G.~D.}\ \bibnamefont{Cohen}},\ }%
  \bibfield{journal}{%
  \bibinfo {journal} {J. Stat. Phys.}\ }%
  \textbf{\bibinfo {volume} {80}},\ \bibinfo {pages} {931} (\bibinfo {year}
  {1995})%
  \bibAnnoteFile{NoStop}{Gallavotti95b}%
\bibitem{SearlesEvans00}%
  \BibitemOpen
  \bibfield{author}{%
  \bibinfo {author} {\bibfnamefont{D.~J.}\ \bibnamefont{Searles}}\ and\
  \bibinfo {author} {\bibfnamefont{D.~J.}\ \bibnamefont{Evans}},\ }%
  \bibfield{journal}{%
  \bibinfo {journal} {J. Chem. Phys.}\ }%
  \textbf{\bibinfo {volume} {113}},\ \bibinfo {pages} {3503} (\bibinfo {year}
  {2000})%
  \bibAnnoteFile{NoStop}{SearlesEvans00}%
\bibitem{EvansMorrisB}%
  \BibitemOpen
  \bibfield{author}{%
  \bibinfo {author} {\bibfnamefont{D.~J.}\ \bibnamefont{Evans}}\ and\ \bibinfo
  {author} {\bibfnamefont{G.}~\bibnamefont{Morris}},\ }%
  \emph{\bibinfo {title} {Statistical Mechanics of Nonequilibrium Liquids}},\
  \bibinfo {edition} {2nd}\ ed.\ (\bibinfo {publisher} {Cambridge University
  Press},\ \bibinfo {year} {2008})%
  \bibAnnoteFile{NoStop}{EvansMorrisB}%
\bibitem{Kurchan98}%
  \BibitemOpen
  \bibfield{author}{%
  \bibinfo {author} {\bibfnamefont{J.}~\bibnamefont{Kurchan}},\ }%
  \bibfield{journal}{%
  \bibinfo {journal} {J. Phys. A}\ }%
  \textbf{\bibinfo {volume} {31}},\ \bibinfo {pages} {3719} (\bibinfo {year}
  {1998})%
  \bibAnnoteFile{NoStop}{Kurchan98}%
\bibitem{Lebowitz99}%
  \BibitemOpen
  \bibfield{author}{%
  \bibinfo {author} {\bibfnamefont{J.~L.}\ \bibnamefont{Lebowitz}}\ and\
  \bibinfo {author} {\bibfnamefont{H.}~\bibnamefont{Spohn}},\ }%
  \bibfield{journal}{%
  \bibinfo {journal} {J. Stat. Phys.}\ }%
  \textbf{\bibinfo {volume} {95}},\ \bibinfo {pages} {333} (\bibinfo {year}
  {1999})%
  \bibAnnoteFile{NoStop}{Lebowitz99}%
\bibitem{Crooks99}%
  \BibitemOpen
  \bibfield{author}{%
  \bibinfo {author} {\bibfnamefont{G.~E.}\ \bibnamefont{Crooks}},\ }%
  \bibfield{journal}{%
  \bibinfo {journal} {Phys. Rev. E}\ }%
  \textbf{\bibinfo {volume} {60}},\ \bibinfo {pages} {2721} (\bibinfo {year}
  {1999})%
  \bibAnnoteFile{NoStop}{Crooks99}%
\bibitem{Crooks00}%
  \BibitemOpen
  \bibfield{author}{%
  \bibinfo {author} {\bibfnamefont{G.~E.}\ \bibnamefont{Crooks}},\ }%
  \bibfield{journal}{%
  \bibinfo {journal} {Phys. Rev. E}\ }%
  \textbf{\bibinfo {volume} {61}},\ \bibinfo {pages} {2361} (\bibinfo {year}
  {2000})%
  \bibAnnoteFile{NoStop}{Crooks00}%
\bibitem{Seifert05}%
  \BibitemOpen
  \bibfield{author}{%
  \bibinfo {author} {\bibfnamefont{U.}~\bibnamefont{Seifert}},\ }%
  \bibfield{journal}{%
  \bibinfo {journal} {Phys. Rev. Lett.}\ }%
  \textbf{\bibinfo {volume} {95}},\ \bibinfo {pages} {040602} (\bibinfo {year}
  {2005})%
  \bibAnnoteFile{NoStop}{Seifert05}%
\bibitem{EspositoVdBPRL10}%
  \BibitemOpen
  \bibfield{author}{%
  \bibinfo {author} {\bibfnamefont{M.}~\bibnamefont{Esposito}}\ and\ \bibinfo
  {author} {\bibfnamefont{C.}~\bibnamefont{Van~den Broeck}},\ }%
  \bibfield{journal}{%
  \bibinfo {journal} {Phys. Rev. Lett.}\ }%
  \textbf{\bibinfo {volume} {104}},\ \bibinfo {pages} {090601} (\bibinfo {year}
  {2010})%
  \bibAnnoteFile{NoStop}{EspositoVdBPRL10}%
\bibitem{Jarzynski97}%
  \BibitemOpen
  \bibfield{author}{%
  \bibinfo {author} {\bibfnamefont{C.}~\bibnamefont{Jarzynski}},\ }%
  \bibfield{journal}{%
  \bibinfo {journal} {Phys. Rev. Lett.}\ }%
  \textbf{\bibinfo {volume} {78}},\ \bibinfo {pages} {2690} (\bibinfo {year}
  {1997})%
  \bibAnnoteFile{NoStop}{Jarzynski97}%
\bibitem{Jarzynski97b}%
  \BibitemOpen
  \bibfield{author}{%
  \bibinfo {author} {\bibfnamefont{C.}~\bibnamefont{Jarzynski}},\ }%
  \bibfield{journal}{%
  \bibinfo {journal} {Phys. Rev. E}\ }%
  \textbf{\bibinfo {volume} {56}},\ \bibinfo {pages} {5018} (\bibinfo {year}
  {1997})%
  \bibAnnoteFile{NoStop}{Jarzynski97b}%
\bibitem{HTasaki00}%
  \BibitemOpen
  \bibfield{author}{%
  \bibinfo {author} {\bibfnamefont{H.}~\bibnamefont{Tasaki}},\ }%
  \bibfield{journal}{%
  \bibinfo {journal} {cond-mat/0009244}}%
   (\bibinfo {year} {2000})%
  \bibAnnoteFile{NoStop}{HTasaki00}%
\bibitem{Maes03}%
  \BibitemOpen
  \bibfield{author}{%
  \bibinfo {author} {\bibfnamefont{C.}~\bibnamefont{Maes}},\ }%
  \bibfield{journal}{%
  \bibinfo {journal} {S\'eminaire Poincar\'e}\ }%
  \textbf{\bibinfo {volume} {2}},\ \bibinfo {pages} {29} (\bibinfo {year}
  {2003})%
  \bibAnnoteFile{NoStop}{Maes03}%
\bibitem{Mukamel03}%
  \BibitemOpen
  \bibfield{author}{%
  \bibinfo {author} {\bibfnamefont{S.}~\bibnamefont{Mukamel}},\ }%
  \bibfield{journal}{%
  \bibinfo {journal} {Phys. Rev. Lett.}\ }%
  \textbf{\bibinfo {volume} {90}},\ \bibinfo {pages} {170604} (\bibinfo {year}
  {2003})%
  \bibAnnoteFile{NoStop}{Mukamel03}%
\bibitem{Monnai05}%
  \BibitemOpen
  \bibfield{author}{%
  \bibinfo {author} {\bibfnamefont{T.}~\bibnamefont{Monnai}},\ }%
  \bibfield{journal}{%
  \bibinfo {journal} {Phys. Rev. E}\ }%
  \textbf{\bibinfo {volume} {72}},\ \bibinfo {pages} {027102} (\bibinfo {year}
  {2005})%
  \bibAnnoteFile{NoStop}{Monnai05}%
\bibitem{EspositoReview}%
  \BibitemOpen
  \bibfield{author}{%
  \bibinfo {author} {\bibfnamefont{M.}~\bibnamefont{Esposito}}, \bibinfo
  {author} {\bibfnamefont{U.}~\bibnamefont{Harbola}},\ and\ \bibinfo {author}
  {\bibfnamefont{S.}~\bibnamefont{Mukamel}},\ }%
  \bibfield{journal}{%
  \bibinfo {journal} {Rev. Mod. Phys.}\ }%
  \textbf{\bibinfo {volume} {81}},\ \bibinfo {pages} {1665} (\bibinfo {year}
  {2009})%
  \bibAnnoteFile{NoStop}{EspositoReview}%
\bibitem{AndrieuxGaspardTasaki}%
  \BibitemOpen
  \bibfield{author}{%
  \bibinfo {author} {\bibfnamefont{D.}~\bibnamefont{Andrieux}}, \bibinfo
  {author} {\bibfnamefont{P.}~\bibnamefont{Gaspard}}, \bibinfo {author}
  {\bibfnamefont{T.}~\bibnamefont{Monnai}},\ and\ \bibinfo {author}
  {\bibfnamefont{S.}~\bibnamefont{Tasaki}},\ }%
  \bibfield{journal}{%
  \bibinfo {journal} {New J. Phys.}\ }%
  \textbf{\bibinfo {volume} {11}},\ \bibinfo {pages} {043014} (\bibinfo {year}
  {2009})%
  \bibAnnoteFile{NoStop}{AndrieuxGaspardTasaki}%
\bibitem{Crooks98}%
  \BibitemOpen
  \bibfield{author}{%
  \bibinfo {author} {\bibfnamefont{G.~E.}\ \bibnamefont{Crooks}},\ }%
  \bibfield{journal}{%
  \bibinfo {journal} {J. Stat. Phys.}\ }%
  \textbf{\bibinfo {volume} {90}},\ \bibinfo {pages} {1481} (\bibinfo {year}
  {1998})%
  \bibAnnoteFile{NoStop}{Crooks98}%
\bibitem{SeifertST08}%
  \BibitemOpen
  \bibfield{author}{%
  \bibinfo {author} {\bibfnamefont{U.}~\bibnamefont{Seifert}},\ }%
  \bibfield{journal}{%
  \bibinfo {journal} {Eur. Phys. J. B}\ }%
  \textbf{\bibinfo {volume} {64}},\ \bibinfo {pages} {423} (\bibinfo {year}
  {2008})%
  \bibAnnoteFile{NoStop}{SeifertST08}%
\bibitem{Harris07}%
  \BibitemOpen
  \bibfield{author}{%
  \bibinfo {author} {\bibfnamefont{R.~J.}\ \bibnamefont{Harris}}\ and\ \bibinfo
  {author} {\bibfnamefont{G.~M.}\ \bibnamefont{Schutz}},\ }%
  \bibfield{journal}{%
  \bibinfo {journal} {J. Stat. Mech.},\ \bibinfo {pages} {P07020}}%
   (\bibinfo {year} {2007})%
  \bibAnnoteFile{NoStop}{Harris07}%
\bibitem{EspoVdB10_Da}%
  \BibitemOpen
  \bibfield{author}{%
  \bibinfo {author} {\bibfnamefont{M.}~\bibnamefont{Esposito}}\ and\ \bibinfo
  {author} {\bibfnamefont{C.}~\bibnamefont{Van~den Broeck}},\ }%
  \bibinfo {journal} {1005.1683}%
  \bibAnnoteFile{NoStop}{EspoVdB10_Da}%
\bibitem{EspoVdB10_Db}%
  \BibitemOpen
\bibfield{journal}{%
    }%
  \bibfield{author}{%
  \bibinfo {author} {\bibfnamefont{C.}~\bibnamefont{Van~den Broeck}}\ and\
  \bibinfo {author} {\bibfnamefont{M.}~\bibnamefont{Esposito}},\ }%
  \bibinfo {journal} {1005.1686}%
  \bibAnnoteFile{NoStop}{EspoVdB10_Db}%
\bibitem{VandenBroeck07}%
  \BibitemOpen
\bibfield{journal}{%
    }%
  \bibfield{author}{%
  \bibinfo {author} {\bibfnamefont{R.}~\bibnamefont{Kawai}}, \bibinfo {author}
  {\bibfnamefont{J.~M.~R.}\ \bibnamefont{Parrondo}},\ and\ \bibinfo {author}
  {\bibfnamefont{C.}~\bibnamefont{Van~den Broeck}},\ }%
  \bibfield{journal}{%
  \bibinfo {journal} {Phys. Rev. Lett.}\ }%
  \textbf{\bibinfo {volume} {98}},\ \bibinfo {pages} {080602} (\bibinfo {year}
  {2007})%
  \bibAnnoteFile{NoStop}{VandenBroeck07}%
\bibitem{Parrondo2009}%
  \BibitemOpen
  \bibfield{author}{%
  \bibinfo {author} {\bibfnamefont{J.~M.~R.}\ \bibnamefont{Parrondo}}, \bibinfo
  {author} {\bibfnamefont{C.}~\bibnamefont{Van~den Broeck}},\ and\ \bibinfo
  {author} {\bibfnamefont{R.}~\bibnamefont{Kawai}},\ }%
  \bibfield{journal}{%
  \bibinfo {journal} {New Journal of Physics}\ }%
  \textbf{\bibinfo {volume} {11}},\ \bibinfo {pages} {073008} (\bibinfo {year}
  {2009})%
  \bibAnnoteFile{NoStop}{Parrondo2009}%
\bibitem{JarzynskiEPL09}%
  \BibitemOpen
  \bibfield{author}{%
  \bibinfo {author} {\bibfnamefont{S.}~\bibnamefont{Vaikuntanathan}}\ and\
  \bibinfo {author} {\bibfnamefont{C.}~\bibnamefont{Jarzynski}},\ }%
  \bibfield{journal}{%
  \bibinfo {journal} {EPL}\ }%
  \textbf{\bibinfo {volume} {87}},\ \bibinfo {pages} {60005} (\bibinfo {year}
  {2009})%
  \bibAnnoteFile{NoStop}{JarzynskiEPL09}%
\bibitem{EspoLindVdBNJP10}%
  \BibitemOpen
  \bibfield{author}{%
  \bibinfo {author} {\bibfnamefont{M.}~\bibnamefont{Esposito}}, \bibinfo
  {author} {\bibfnamefont{K.}~\bibnamefont{Lindenberg}},\ and\ \bibinfo
  {author} {\bibfnamefont{C.}~\bibnamefont{Van~den Broeck}},\ }%
  \bibfield{journal}{%
  \bibinfo {journal} {New J. Phys.}\ }%
  \textbf{\bibinfo {volume} {12}},\ \bibinfo {pages} {013013} (\bibinfo {year}
  {2010})%
  \bibAnnoteFile{NoStop}{EspoLindVdBNJP10}%
\bibitem{Gallavotti08}%
  \BibitemOpen
  \bibfield{author}{%
  \bibinfo {author} {\bibfnamefont{G.}~\bibnamefont{Gallavotti}},\ }%
  \bibfield{journal}{%
  \bibinfo {journal} {Eur. Phys. J. B}\ }%
  \textbf{\bibinfo {volume} {61}},\ \bibinfo {pages} {1} (\bibinfo {year}
  {2008})%
  \bibAnnoteFile{NoStop}{Gallavotti08}%
\bibitem{Ruelle03}%
  \BibitemOpen
  \bibfield{author}{%
  \bibinfo {author} {\bibfnamefont{D.}~\bibnamefont{Ruelle}},\ }%
  \bibfield{journal}{%
  \bibinfo {journal} {PNAS}\ }%
  \textbf{\bibinfo {volume} {100}},\ \bibinfo {pages} {3054} (\bibinfo {year}
  {2003})%
  \bibAnnoteFile{NoStop}{Ruelle03}%
\bibitem{Holian}%
  \BibitemOpen
  \bibfield{author}{%
  \bibinfo {author} {\bibfnamefont{B.~L.}\ \bibnamefont{Holian}},\ }%
  \bibfield{journal}{%
  \bibinfo {journal} {Phys. Rev. A}\ }%
  \textbf{\bibinfo {volume} {33}},\ \bibinfo {pages} {1152} (\bibinfo {year}
  {1986})%
  \bibAnnoteFile{NoStop}{Holian}%
\bibitem{Nose84}%
  \BibitemOpen
  \bibfield{author}{%
  \bibinfo {author} {\bibfnamefont{S.}~\bibnamefont{Nose}},\ }%
  \bibfield{journal}{%
  \bibinfo {journal} {Molecular Physics}\ }%
  \textbf{\bibinfo {volume} {52}},\ \bibinfo {pages} {255} (\bibinfo {year}
  {1984})%
  \bibAnnoteFile{NoStop}{Nose84}%
\bibitem{Hoover85}%
  \BibitemOpen
  \bibfield{author}{%
  \bibinfo {author} {\bibfnamefont{W.~G.}\ \bibnamefont{Hoover}},\ }%
  \bibfield{journal}{%
  \bibinfo {journal} {Phys. Rev. A}\ }%
  \textbf{\bibinfo {volume} {31}},\ \bibinfo {pages} {1695} (\bibinfo {year}
  {1985})%
  \bibAnnoteFile{NoStop}{Hoover85}%
\bibitem{CoverThomas}%
  \BibitemOpen
  \bibfield{author}{%
  \bibinfo {author} {\bibfnamefont{T.~M.}\ \bibnamefont{Cover}}\ and\ \bibinfo
  {author} {\bibfnamefont{J.~A.}\ \bibnamefont{Thomas}},\ }%
  \emph{\bibinfo {title} {Elements of information theory}}\ (\bibinfo
  {publisher} {Wiley},\ \bibinfo {year} {2006})%
  \bibAnnoteFile{NoStop}{CoverThomas}%
\bibitem{Evans85}%
  \BibitemOpen
  \bibfield{author}{%
  \bibinfo {author} {\bibfnamefont{D.~J.}\ \bibnamefont{Evans}},\ }%
  \bibfield{journal}{%
  \bibinfo {journal} {Phys. Rev. A}\ }%
  \textbf{\bibinfo {volume} {32}},\ \bibinfo {pages} {2923} (\bibinfo {year}
  {1985})%
  \bibAnnoteFile{NoStop}{Evans85}%
\bibitem{Gaspard04b}%
  \BibitemOpen
  \bibfield{author}{%
  \bibinfo {author} {\bibfnamefont{P.}~\bibnamefont{Gaspard}},\ }%
  \bibfield{journal}{%
  \bibinfo {journal} {J. Stat. Phys}\ }%
  \textbf{\bibinfo {volume} {117}},\ \bibinfo {pages} {599} (\bibinfo {year}
  {2004})%
  \bibAnnoteFile{NoStop}{Gaspard04b}%
\end{thebibliography}

%

\end{document}